\documentclass[prl,amssymb,twocolumn,superscriptaddress,showpacs,floatfix]{revtex4}

\usepackage{graphicx}
\usepackage{dcolumn}
\usepackage{bm}

\begin{document}

\title{Validation of Lasing in Active Nanocavities}

\author{Yong-Seok Choi} \email[Corresponding author: ]{cys@cnsi.ucsb.edu}
\affiliation{ECE Department, University of California Santa
Barbara, CA 93106, Santa Barbara, USA}

\author{Matthew T. Rakher}
\affiliation{Department of Physics, University of California Santa
Barbara, CA 93106, Santa Barbara, USA}

\author{Kevin Hennessy}
\affiliation{ECE Department, University of California Santa
Barbara, CA 93106, Santa Barbara, USA}

\author{Stefan Strauf}
\affiliation{Department of Physics, University of California Santa
Barbara, CA 93106, Santa Barbara, USA} \affiliation{Materials
Department, University of California Santa Barbara, CA 93106,
Santa Barbara, USA}

\author{Antonio Badolato}
\affiliation{ECE Department, University of California Santa
Barbara, CA 93106, Santa Barbara, USA}

\author{Pierre M. Petroff}
\affiliation{ECE Department, University of California Santa
Barbara, CA 93106, Santa Barbara, USA} \affiliation{Materials
Department, University of California Santa Barbara, CA 93106,
Santa Barbara, USA}

\author{Dirk Bouwmeester}
\affiliation{Department of Physics, University of California Santa
Barbara, CA 93106, Santa Barbara, USA}

\author{Evelyn L. Hu}
\affiliation{ECE Department, University of California Santa
Barbara, CA 93106, Santa Barbara, USA} \affiliation{Materials
Department, University of California Santa Barbara, CA 93106,
Santa Barbara, USA}

\date{\today}

\begin{abstract}
An unambiguous proof of lasing in an active nanocavity with
ultrahigh spontaneous emission coupling factor ($\beta$ = 0.65) is
presented. To distinguish the subtle lasing threshold features
from possible material-related phenomena, such as saturable
absorption in the gain medium, a series of active nanocavities
with different values of $\beta$ have been designed to
systematically approach the high-$\beta$ device. The demonstration
of the lasing threshold is obtained through the observation of the
transition from thermal to coherent light photon statistics that
is well understood and identified in the $\beta$ $\ll$ 1 lasing
regime. The systematic investigation allows a more definitive
validation of the onset of lasing in these active nanocavities.
\end{abstract}


\maketitle

Advances in fabrication of optical nano-scale devices enable the
construction of nanolasers in which there is effectively only one
optical emission mode. The $\beta$ factor is a measure inversely
proportional to the number of available modes in which the gain
medium can spontaneously emit photons. Achieving a high $\beta$
factor is a key issue for improving single-photon sources
\cite{Michler-Science2000,Santori-Nature2002,Yuan-Science2002}. In
addition, high $\beta$ values enable low-threshold lasers with
high modulation speeds \cite{Yokoyama} and low intensity noise
\cite{Bjork-APL1994}. Various semiconductor laser systems have
demonstrated high $\beta$ values of order of 0.1
\cite{Slusher,Baba-QE1997,Painter-Science1999,Park-Science2004},
and in particular, recent achievements in quantum dot nanocavities
have demonstrated close-to-perfect coupling efficiencies
\cite{Pelton-PRL2002,Kevin,Finley-PRB05,Stefan}. For such devices
with $\beta$ approaching unity, the lasing threshold is harder to
determine and shifts to very low optical pump and output powers
(the hypothetical case of a $\beta$ = 1 is often referred to as a
thresholdless laser \cite{Yokoyama}). The characteristic
nonlinearities in the optical output power and the emission
linewidth around the lasing threshold become so subtle that one
might wonder whether other effects such as saturable absorption in
the optical gain medium could be responsible for the observed
features \cite{Nick-APL05}. The question of what constitutes a
laser becomes even more intricate if an unconventional gain
medium, e.g. consisting of a very low density of optically active
quantum dots, is considered.

In this Letter, we provide validation of the existence of a lasing
threshold in a $\beta$ factor of nearly unity laser that is
independent of the precise gain mechanism and the (nonlinear)
absorption characteristics of the material. The validation is
based on the design and fabrication of a series of nanocavities
that support dominantly, one, two and three optical modes
overlapping with the gain medium. This stepwise decreases the
$\beta$ factor from approximately 1 to 0.5 to 0.33 (in our
experiments from 0.65 to 0.35 to 0.15) and brings the observed
transition features into the regime where they are unambiguously
identified as the lasing threshold. As the defining signature of
the lasing threshold, we use the transition from the thermal light
source to the coherent light source as measured by the
second-order intensity correlation function:
\begin{equation} g^{(2)}(\tau )=\frac{\langle
{E}^{(-)}(t){E}^{(-)}(t+\tau){E}^{(+)}(t+\tau){E}^{(+)}(t) \rangle
} {(\langle {E}^{(-)}(t){E}^{(+)}(t) \rangle)^2}
\end{equation}
where $E^{(+)}(t)$ and $E^{(-)}(t)$ are the positive and negative
frequency parts of the electromagnetic field, respectively
\cite{Glauber}. The measurement of $g^{(2)}(\tau)$ reveals a
fundamental difference between a laser, i.e. $g^{(2)}(\tau)$ = 1,
and thermal light, i.e. $g^{(2)}(0)$ = 2 \cite{Arecchi-PL1966}.
The interpretation of the observed-laser-photon statistics
\cite{Freed-PRL1965,Smith-PRL1966} led to the quantum theory on
coherent-state transitions \cite{Scully,Filipowicz-PRA1986}.

\begin{figure}
\includegraphics{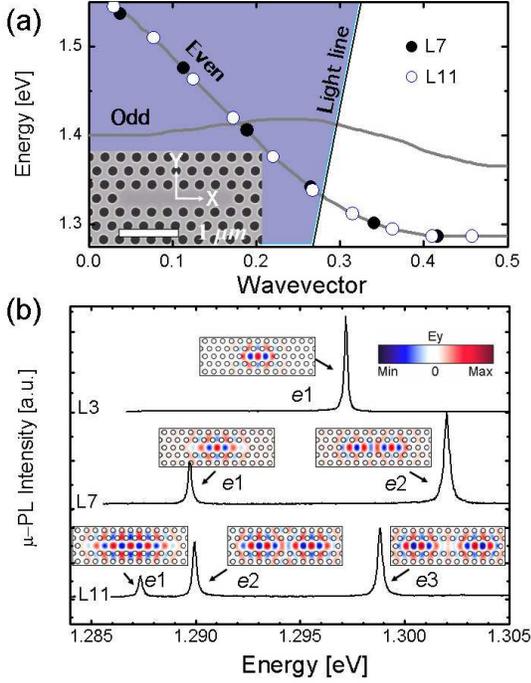}
\caption{\label{fig:LnCavity} (a) The dispersion curves of a
single-line-defect PC waveguide and the resonances of the L7 and
L11 cavities obtained by 3D FDTD, where the shaded region is a
leaky region. An inset shows the SEM of an L7 cavity. (b)
Calculated $E_y$ field profiles identify resonant peaks in the
$\mu$-PL spectra.}
\end{figure}

We consider nanocavities with a gain medium comprised of a low
density ($\sim$ 5$\times$10$^9$/cm$^2$) of InAs quantum dots (QDs)
grown by a partially covered island technique \cite{Garcia-APL98}.
Nanocavities were designed to support one, two, and three modes in
the s-shell region of QDs, and hence to reveal the effect of
$\beta$ on laser photon statistics systematically. Nanocavities
utilize the triangular-lattice photonic-crystal (PC) structures
with three, seven, and eleven missing holes in the $\Gamma$-$K$
direction, denoted as L3, L7, and L11 cavities, respectively. The
conditions for cavity resonances are determined by the PC
waveguide dispersion and the Fabry-Perot condition
\cite{Kim-JAP2004}. FIG. \ref{fig:LnCavity}(a) shows the
dispersion curves of a single-line-defect PC waveguide and the
resonant modes of L7 and L11 cavities calculated by the
three-dimensional (3D) finite-difference time-domain (FDTD)
method. To use the low energy, high-$Q$ even (e\#) modes below the
light line, we specified a lattice constant of $\sim$ 260 nm, a
hole radius of $\sim$ 65 nm, and a membrane thickness of $\sim$
126 nm in the processing \cite{Kevin}. An L7 cavity is shown in
the scanning electron micrograph (SEM) inset of FIG.
\ref{fig:LnCavity}(a).

The experimental setup consists of a He-cryostat,
micro-photoluminescence ($\mu$-PL) system, and a photon
correlation measurement apparatus with two avalanche photodiodes
(APDs) in the Hanbury-Brown Twiss configuration. The QD PC
nanolasers were optically pumped by a power and temperature
stabilized 780 nm diode laser. FIG. \ref{fig:LnCavity}(b) shows
the $\mu$-PL spectra of L3, L7, and L11 cavities with one, two,
and three modes, respectively. All the modes are polarized in the
Y direction with no degeneracy. Provided that these nanocavity
modes are decorated by the broad QD gain spectrum \cite{Stefan},
the maximum $\beta$ is 1 for L3, 0.5 for L7, and 0.33 for L11
cavities.

\begin{figure}
\includegraphics{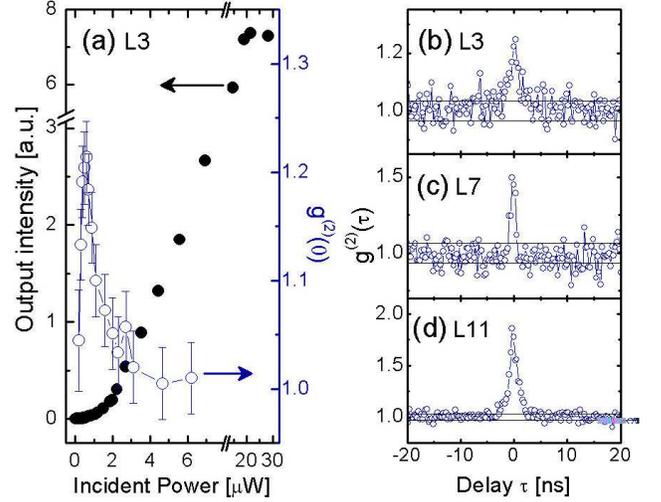}
\caption{\label{fig:L3L7L11} (a) The output intensity and the
$g^{(2)}(0)$ graph of an L3 laser as a function of the incident
power. The $g^{(2)}(\tau)$ spectra with maximum bunching behaviors
observed in (b) L3, (c) L7, and (d) L11 lasers, respectively. In
the $g^{(2)}(\tau)$ measurements, the spectral window was set by
the 0.5 nm bandpass filters attached to the APDs. The APD
detection rates were kept to be 10$^4$ - 10$^5$ at various pump
powers. A coincidence value at long delay time ($\tau$ $>$ 100 ns)
was taken as a reference for normalization. Parallel lines in
(b)-(d) indicate the standard deviation noise.}
\end{figure}

To explore the lasing behavior of the QD PC nanocavities, we can
inspect a typical output characteristics of an L3 laser, plotted
as a function of the incident pump power as shown in FIG.
\ref{fig:L3L7L11}(a). The output starts to grow nonlinearly, and
then increases in a nearly linear manner before being clamped at
values of incident pump powers above 20 $\mu$W. The dependence of
$g^{(2)}(0)$ on the incident pump power is compared with the
output characteristics in FIG. \ref{fig:L3L7L11}(a). $g^{(2)}(0)$
increases with incident power from 0.2 $\mu$W to 0.6 $\mu$W,
subsequently converging to one before the output emission is
clamped. FIG. \ref{fig:L3L7L11}(b) is the $g^{(2)}(\tau)$ plot
taken at 0.6 $\mu$W, which shows the largest bunching peak for the
L3 laser. This maximum bunching plot for the L3 structure can be
compared with the similar plots for the L7 and L11 lasers, shown
in FIGs. \ref{fig:L3L7L11}(c) and \ref{fig:L3L7L11}(d). Note that
there are systematic variations in the peak value of
$g^{(2)}(\tau)$, the width of the bunching peak, and the signal to
noise of the curves. These variations are the natural outcome of
the different values of $\beta$, determined by the number of
allowed cavity modes in the different structures, as we will show
below. One immediate question raised by the data of FIG.
\ref{fig:L3L7L11}(a) is the decrease of $g^{(2)}(0)$ below
threshold, as it should be super-Poissonian with $g^{(2)}(0)$ = 2
\cite{Arecchi-PL1966}.

\begin{figure}[tb!]
\includegraphics[width=85mm]{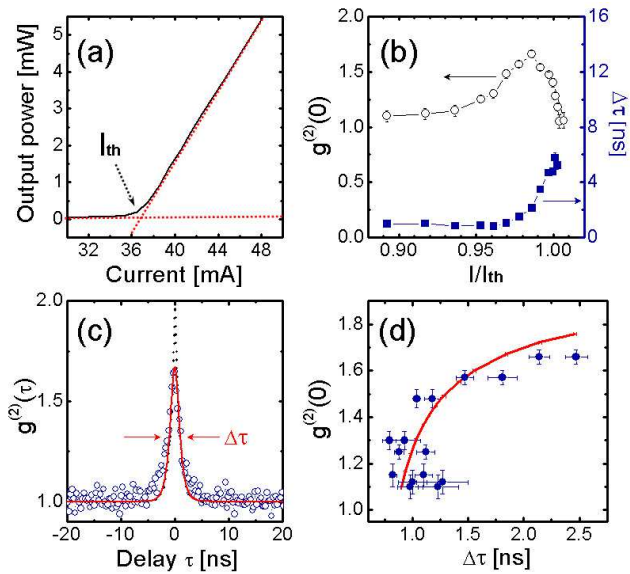}
\caption{\label{fig:QWLD} (a) The output power of the MQW laser as
a function of current $I$. (b) $g^{(2)}(0)$ and $\Delta \tau$ as a
function of $I/I_{th}$. (c) The experimental (open circles) and
theoretical $g^{(2)}(\tau)$ spectra, where the dashed line is the
ideal $g^{(2)}(\tau)$ while the solid line is the simulated
$g^{(2)}(\tau)$ with $M$($t$,$\sigma=$450ps). (d) The effect of
$M(t,\sigma)$ on $g^{(2)}(0)$ and $\Delta \tau$, where dots are
the experimental data and the line is a simulation.}
\end{figure}

To gain further understanding into this discrepancy, we sought
insight through calibration measurements of a more conventional
GaAlAs MQW laser that also serves as the pump source for our QD PC
nanolasers. The laser displays a clear threshold current of 36.25
mA as shown in FIG. \ref{fig:QWLD}(a). Below threshold, the MQW
gain spectrum decorates about $100$ cavity modes with a mode
spacing of 0.13 nm. To discriminate side modes in the photon
correlation measurements, we replaced the 0.5 nm bandpass filters
by a monochromator with a bandwidth as narrow as the lasing mode
linewidth. Pronounced bunching signals in $g^{(2)}(\tau)$ were
observed in the narrow current region from $0.88\times I_{th}$ to
$1.01\times I_{th}$. The bunching height, i.e. $g^{(2)}(0)$,
increases as the current is elevated to $0.986 \times I_{th}$ and
rapidly converges to one as shown in FIG. \ref{fig:QWLD}(b). The
bunching width, i.e. $\Delta \tau$, is nearly constant ($\sim$ 1
ns) with operating currents from $0.88 \times I_{th}$ to $0.97
\times I_{th}$ where $g^{(2)}(0)$ changes from 1.1 to 1.5. Then,
$\Delta \tau$ grows with the increase of operating current as
shown in FIG. \ref{fig:QWLD}(b). The observed behaviors of
$g^{(2)}(0)$ and $\Delta \tau$ over threshold are expected as the
emission becomes more coherent. But, the absolute value of
$g^{(2)}(0)$ below threshold is naively expected to obtain the
value of 2 that does not match the data. This can be explained by
taking into account the limited temporal resolution of 450 ps of
our setup. When the coherence time of the light source becomes
comparable to the temporal resolution of the measurement system,
then the $g^{(2)}(\tau)$ below threshold, which is normally given
by $g^{(2)}(\tau)$ $=$ $1 + {\rm exp} [-2|\tau |/\tau_c]$, must be
convolved with a measurement function $M(t,\sigma)$ $=$ $(\sigma
\sqrt{2 \pi})^{-1} {\rm exp} [-t^2 / 2 \sigma^2]$, where $\sigma$
is the APD timing jitter. FIG. \ref{fig:QWLD}(c) shows the ideal
and simulated $g^{(2)}(\tau)$ functions for $\tau_c$ = 2 ns and
$\sigma$ = 450 ps. The simulation shows that $g^{(2)}(0 )$ is
reduced by 0.3, in good agreement with the experimental data taken
at 0.986 $\times$ $I_{th}$. We examined the effect of
$M(\tau,\sigma)$ for various values of $\tau_c$ from 0.1 ns to 2
ns, and the results are plotted with the experimental data in FIG.
\ref{fig:QWLD}(d), showing excellent agreement. The calibration
measurements underscore the importance of adequate timing
resolution for sources with short coherence times. These
measurements also demonstrate that the genuine signature of lasing
action is the convergence of $g^{(2)}(0)$ to one above threshold.

To complete our quantitative analysis of the variation of
$g^{(2)}(0)$ with pump power, the identification of the onset of
lasing, and the correlations among QD PC nanolasers with different
$\beta$ values, we first calculate the photon number probability
($p_n$), given by \cite{Filipowicz-PRA1986,Jin-PRA1994},
\begin{equation}
p_n = p_0 \prod_{k=1}^{k=n} \frac{N_a}{N_b+2T_1 (R^{-1}+k)\kappa},
\end{equation}
where $p_0$ is the zero photon probability and can be determined
by normalization, $n$ is the number of photons in the cavity, and
$N_a$ and $N_b$ are the number of carriers in the upper and the
lower levels, respectively. $T_1$ is defined by
$T_1=(\Lambda_0+\gamma')^{-1}$, where $\Lambda_0$ is the pumping
rate and $\gamma'$ is the sum of all recombination rates that do
not add a photon to the cavity mode. $R$ is defined by
$R=4g^2T_1T_2$, where $g$ is the coupling parameter between the
cavity mode and the excitonic transitions of carriers, and $T_2$
is the dephasing time. $\kappa$ is the cavity decay rate given by
$\kappa=2\pi \nu /Q$, where $\nu$ is the frequency of the cavity
mode. The normalized zero-field populations of the upper and lower
laser levels are given by $N_a = N \Lambda_0 T_1$ and $N_b= N
\gamma'T_1$, where $N$ is the total number of carriers that
correspond to gain saturation.

To investigate the effect of $\beta$ on the laser photon
statistics, we adopted the phenomenological expression
\cite{Davidovich-PRA1999},
\begin{equation}
\beta = \left [ 1+ \frac{\gamma'}{2g^2} \left (
\frac{1}{T_2}+\frac{\kappa}{2} \right ) \right ]^{-1},
\end{equation}
which is simply the ratio of the spontaneous emission rate into
the cavity mode to the total recombination rate. We used $\beta$
and $N$ as fitting parameters to best describe the observed
${g^{(2)}(0)}\rightarrow 1$ behavior with the increase of the pump
power. Here, $N$ rather than the number of QDs was introduced as a
fitting parameter to account for the non-resonant optical pumping
scheme. In contrast to the resonant optical excitation of QD
energy levels, photo-generated carriers in the GaAs bulk region
relax to the wetting layer (WL) and to QDs. In this process, the
efficient coupling between the rich WL density of states and QDs
is believed to be fast enough to repopulate QDs contributing to
lasing action \cite{Stefan}. For the QD PC nanolasers at 4 K,
$\gamma'$ can be obtained by $2\pi/\tau_{PL}$ according to the
$\mu$-PL decay time ($\tau_{PL} $$\sim$ 5 ns) measured for a QD
uncoupled to the cavity mode. The theoretical analysis was
unaffected by taking $T_2$ in the range of 0.1 ns to 1 ns
\cite{Santori-Nature2002,Borri-PRL2001}. For the GaAlAs MQW laser
at the room temperature, we can use the typical parameters of
$\gamma'$ of 1.0 GHz and $T_2$ of 0.1 ps and $N$ of
5$\times$10$^7$ \cite{Jin-PRA1994}. Other parameters used in
calculation are summarized in Table \ref{tab:table1}.

\begin{table}
\caption{\label{tab:table1}We summarize the spontaneous emission
coupling factor $\beta$, the cavity decay rate $\kappa$, the
exciton-cavity coupling parameter $g$, and the number of carriers
$N$ for QD PC nanolasers and the conventional MQW laser.}
\begin{ruledtabular}
\begin{tabular}{cccccccc}
Device &$\beta$ &$\kappa$ [GHz] &$g$ [GHz] &$N$ \\
\hline
L3 &$0.65\pm0.05$ &287 &$13\mp2$ &$(3.2\mp0.7) \times 10^2$ \\
L7 &$0.35\pm0.05$ &432 &$8.6\mp1.0$ &$(1.1\mp0.2) \times 10^3$ \\
L11 &$0.15\pm0.025$ &396 &$4.7\mp0.5$ &$(3.3\mp0.5) \times 10^3$ \\
MQW laser &$0.0001$ &200 &0.7 &$5 \times 10^7$ \cite{Jin-PRA1994}
\end{tabular}
\end{ruledtabular}
\end{table}

\begin{figure}
\includegraphics{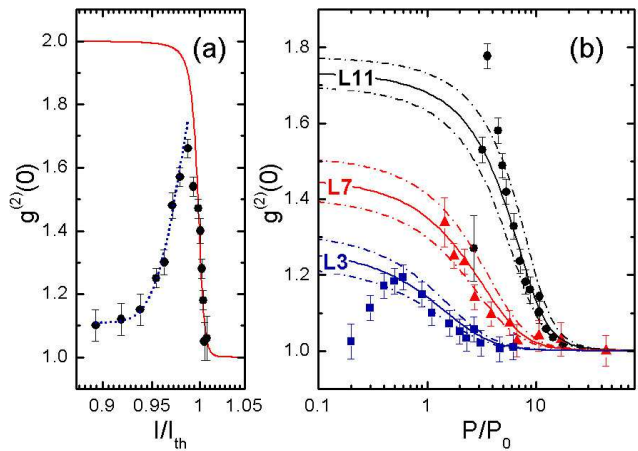}
\caption{\label{fig:FitExp} (a) The experimental $g^{(2)}(0)$ data
(dot) of the MQW laser with the theoretical curve (solid line)
obtained by $\beta=0.0001$. The dotted line shows the effect of a
temporal resolution on the measurement of $g^{(2)}(0)$, when the
ideal $g^{(2)}(0)$ is given by the solid line. (b) The
experimental $g^{(2)}(0)$ data of L3, L7, L11 lasers with the
theoretical curves obtained by the $\beta$ of 0.65 $\pm$ 0.05,
0.35 $\pm$ 0.05, and 0.15 $\pm$ 0.025, respectively. The dash-dot
lines indicate the upper and lower bounds of fits considering the
given uncertainties of $\beta$ factors.}
\end{figure}

Based on this analysis, the theoretical $g^{(2)}(0)$ curves were
obtained as shown in FIG. \ref{fig:FitExp}. The change of
$g^{(2)}(0)$ at the vicinity of $I_{th}$ in the MQW laser agrees
with the theoretical curve indicated by a solid line, which was
obtained with $\beta$ of 0.0001 and typical parameters shown in
Table \ref{tab:table1}. Furthermore, the theoretical curve below
threshold shows the expected behavior of $g^{(2)}(0)=2$. If we
simulate the experimental values of $g^{(2)}(0)$ with this
theoretical curve and $M$($t$,$\sigma$=450ps), we can obtain the
dotted line, which is in good agreement with the experimental
data. The theoretical curves for QD PC nanolasers are summarized
in FIG. \ref{fig:FitExp}(b). The systematic variation of the
$g^{(2)}(0)$ values can be seen with the expected $\beta$ values
of 0.65 $\pm$ 0.05, 0.35 $\pm$ 0.05, and 0.15 $\pm$ 0.025 for the
L3, L7, and L11 cavities, respectively. Here, the fits have been
made to best describe the transition of $g^{(2)}(0)$ with the
increase of the pump power, even if the initial bunching behaviors
are yet to be clarified with better temporal resolution and
detection efficiency. But, both the experimental and theoretical
results demonstrate slow coherent-state transitions
\cite{Carmichael-PRA1994}, distinguishing high-$\beta$ QD PC
nanolasers from the conventional MQW laser. The theoretical
$g^{(2)}(0)$ curves below threshold show saturation behaviors that
decrease as $\beta$ increases, which is expected as $\beta$
approaches unity and coherence is observed for all input pump
powers. This can be explained by the fact that high-$\beta$
devices have a significant contribution from spontaneous emission
into the lasing mode even at very low pump powers, which destroys
a thermal field description with $g^{(2)}(0)$ = 2.

In conclusion, we have demonstrated the systematic correlation
between the coherent-state transition and the lasing action in QD
PC nanocavities. The onsets of coherence in soft-turn-on
nanolasers are confirmed by the convergence to Poissonian
statistics, i.e. $g^{(2)}(0)=1$, with the increase of the pump
power. The bunching behavior near threshold is found to subside
with the increase of $\beta$. The quantitative analysis reveals
the very high $\beta$ of 0.65, 0.35, and 0.15, as predicted for
the L3, L7, and L11 cavities, respectively. This general approach
of constructing a series of nano devices with the limiting case as
the device of interest, might prove useful to identify other
features in nano devices that can be difficult to distinguish from
material properties.

This work was funded under DARPA No. 972-01-1-0027, DMEA No.
H94003-04-2-0403, and NSF NIRT No. 0304678.

\end{document}